\documentclass[journal]{IEEEtran}
\usepackage[pdftex]{graphicx}
\usepackage{amsmath,amssymb,ascmac}
\usepackage{algorithm, algorithmic}
\usepackage{soul, balance}
\usepackage{pifont, todonotes}

\ifCLASSOPTIONcompsoc
  \usepackage[caption=false,font=footnotesize,labelfont=sf,textfont=sf]{subfig}
\else
  \usepackage[caption=false,font=footnotesize]{subfig}
\fi

\usepackage{url}
\usepackage{amsthm}
\newtheorem{definition}{Definition}
\def\bvec#1{\mbox{\boldmath $#1$}}

\begin{document}

\title{Immunization of Pruning Attack in DNN Watermarking Using Constant Weight Code}

\author{Minoru Kuribayashi,~\IEEEmembership{Senior Member,~IEEE,}
        Tatsuya Yasui,~\IEEEmembership{Student Member,~IEEE,}
        Asad Malik,
        Nobuo Funabiki,~\IEEEmembership{Member,~IEEE}
\thanks{The authors are with Okayama University, Okayama, Japan. (e-mail: kminoru@okayama-u.ac.jp)}
}


\maketitle

\begin{abstract} 
To ensure protection of the intellectual property rights of DNN models, watermarking techniques have been investigated to insert side-information into the models without seriously degrading the performance of original task.
One of the threats for the DNN watermarking is the pruning attack such that less important neurons in the model are pruned to make it faster and more compact as well as to remove the watermark.
In this study, we investigate a channel coding approach to resist the pruning attack.
As the channel model is completely different from conventional models like digital images, it has been an open problem what kind of encoding method is suitable for DNN watermarking. 
A novel encoding approach by using constant weight codes to immunize the effects of pruning attacks is presented.
To the best of our knowledge, this is the first study that introduces an encoding technique for DNN watermarking to make it robust against pruning attacks.
\end{abstract}


\IEEEpeerreviewmaketitle

\section{Introduction}
Due to the multiple network layers in a DNN model, there are huge amount of parameters known as network weights to be trained for attaining a local minimum. There are many degrees of freedom in the choice of the weight parameters for embedding watermark. The watermark is inserted in such a way that the degradation of the original task's accuracy of the watermarked model is controlled as small as possible.

The first DNN watermarking was presented in \cite{uchida_ICMR17,nagai_IJMIR18} that watermark is embedded into the weight parameters of a Convolutional Neural Network (CNN) model.
The embedding operation is performed in parallel with the training of the model by introducing an embedding loss function so that the weights are updated according to the watermark as well as the supervised dataset.
In \cite{rouhani_ASPLOS19,chen_ICMR19}, the selection of feature vector is refined from the method in \cite{uchida_ICMR17,nagai_IJMIR18}.

It is reported in \cite{Choromanska2015,dauphin14} that
almost all the local minimum have very similar to the global optimum.
Some empirical evidence shows that for deeper or larger models, a local minimum becomes good enough, since their loss values are almost similar.
Considering this characteristic, the watermark is embedded into some sampled weight values in \cite{Kuribayashi_IHMMSEC2021}.
The robustness requirement for DNN watermarking is the capability of recovering the watermark from a perturbed version of a DNN model.
One instance of perturbation is the pruning of the model.
Essentially, to reduce the computational costs for executing DNN models, redundant neurons are pruned without compromising accuracy.
The purpose of pruning is to remove less important weight parameters from the DNN model that contribute to the loss is small.
If watermark signal is embedded into such less important parameters, it is easily removed or modified by the pruning.
Hence, one of the important requirements for the DNN watermarking is the robustness against the pruning attack \cite{finepruning}  while ensuring that the watermarked parameters are relevant for the original task.

From the viewpoint of communication channel, pruning can be regarded as an erasure channel between the transmitter and the receiver of watermark.
Since significant number of symbols are erased over the channel (e.g. more than half of the weight parameters are erased by pruning), the use of erasure correcting codes is not suitable for this channel.
In this study, we encode the watermark by binary constant weight codes~\cite{schalkwijk, Brouwer} to make it robust against the pruning attack.
The number of symbols used as “1” is fixed in the codeword, and it is designed to be as small as possible.
Thus, most of the symbols in the codeword are “0".
At the embedding of such a codeword, we make a constraint by using two thresholds during the training of a DNN model.
For symbols “1", the amplitude is more than a high threshold, while the amplitude for symbol “0" is controlled to be smaller than a low threshold.
Once a pruning attack is executed, the erasure of weight parameters does not affect the symbols “0" because those symbols can be extracted even though the amplitude is small.
On the other hand, the symbols “1" can be detected correctly due to the high amplitude.
Under the assumption that the values of weight parameters follow Gaussian distribution, the secrecy of the embedded watermark is discussed in this paper.

The rest of this paper is organized as follows:
In Section~\ref{sec:pre}, we give some definitions of parameters.
The proposed method is detailed in Section~\ref{sec:prop}, and numerical examples are presented in Section \ref{sec:example}.
Finally, conclusions to this study are drawn in Section~\ref{sec:conclude}, following by some suggestions for future work.

\section{Preliminaries}
\label{sec:pre}
\subsection{Notations}
The notations of parameters in this paper are summarized as follows:
\begin{itemize}
\item $N$: Number of weight parameters in DNN model
\item $L$: Number of selected weight parameters
\item $\bvec{w}=(w_0, w_1, \ldots , w_{L-1}), w_i\in \mathbb{R}$: Selected weight parameters
\item $\bvec{w^\prime}$: Selected weight parameters after pruning attack
\item $R$: Pruning rate
\item $p=\lfloor RN \rfloor$: Number of pruned weight parameters
\item $k$: Bit length of watermark
\item $\bvec{b}=(b_0, b_1, \ldots , b_{k-1}), b_i\in \{0,1\}$: Binary watermark
\item $\bvec{c}=(c_0, c_1, \ldots , c_{L-1}), c_i\in \{0,1\}$: Binary codeword of constant weight code
\item $\alpha=\sum c_i$: Hamming weight of codewords
\item $T_1$, $T_0$: Thresholds for binary classification, where $0 < T_0 < T_1$ 
\item $sort()$: Sort algorithm with ascending order
\item $sgn()$: Sign function
\end{itemize}

\subsection{Definitions}
At the initial phase of the DNN model, the values of all weight parameters are assigned randomly to start training from scratch. Moreover, the multiple combinations of neural networks, there are a huge amount of parameters in the model. Therefore, from the viewpoint of the central limit theorem, the values of the parameters can be regarded as Gaussian distribution even after training the model. So, we came up with the following definitions.
\begin{definition}
\label{def:gauss}
The values of weight parameters $w_i, 0\le i\le N-1$ follow Gaussian distribution with zero mean and a variance $\sigma^2$, ${\cal N}(0, \sigma^2)$.
\end{definition}

After training, the parameters which absolute values are smaller than a threshold are cut-off to zero for the compression of a DNN model.
The threshold is set in such a way that the accuracy of the
model does not decrease significantly. 
\begin{definition}
\label{def:pruning}
For a given rate $0\le R<1$, the pruning attack cut-off the weight values $w_i=0$ if $|w_i| < \tilde{w}_p$ for $0\le i\le N-1$, where  
\begin{equation}
    \tilde{\bvec{w}} = sort(|\bvec{w}|) = sort(|w_0|, |w_1|, \ldots , |w_{N-1}|),
\end{equation}
$sort()$ is a sort algorithm, and
\begin{equation}
  p=\lfloor RN \rfloor.
\end{equation}
\end{definition}

A binary codeword of constant weight codes has a fixed Hamming weight, and its definition is given as follows.
\begin{definition}
\label{def:CWC}
A codeword $\bvec{c}=(c_0, c_1, \ldots , c_{L-1}), c_i\in\{0,1\}$ of constant weight codes satisfies the condition such that
\begin{equation}
    \sum_{i=0}^{L-1}c_i = \alpha,
\end{equation}
where $\alpha$ is fixed constant.
\end{definition}

We also give the following definition to consider the secrecy of watermark.
\begin{definition}
\label{def:scale}
The length $L$ of codeword is much smaller than the total number  $N$ of weight parameters. 
\end{definition}

\subsection{Encoding and Decoding of Constant Weight Code}
The mapping operation from $k$-bit information into a codeword $\bvec{c}$ with weight $\alpha$ and length $L$ has been proposed by Schalkwijk\cite{schalkwijk}. 

Let ${\cal C}(\alpha, L)$ be a set of vector with a constant weight $\alpha$ and length $L$. The procedure to encode watermark
$\bvec{b}$ into a codeword $\bvec{c}\in {\cal C}(\alpha, L)$ is described in Algorithm \ref{alg_encode}.

\begin{algorithm}[ht]
\caption{Encode $\bvec{b}$ into $\bvec{c}$}
\label{alg_encode}
\begin{algorithmic}[1]
 \renewcommand{\algorithmicrequire}{\textbf{Input:}}
 \renewcommand{\algorithmicensure}{\textbf{Output:}}
\REQUIRE $\alpha$, $L$, $\bvec{b}=(b_0, b_1, \ldots
 , b_{k-1})$, $b_t\in\{0, 1\}$
\ENSURE  $\bvec{c}=(c_0, c_1, \ldots , c_{L-1})$, $c_t\in\{0,1\}$
\STATE $\displaystyle B \leftarrow \sum_{t=0}^{k-1}b_t2^t$;
\STATE $\ell \leftarrow \alpha$;
\FOR{$t=0$ \TO $L-1$}
\IF{$B \ge \displaystyle \binom{L-t-1}{\ell}$}
\STATE $c_{L-t-1} = 1$;
\STATE $B \leftarrow B - \displaystyle \binom{L-t-1}{\ell}$;
\STATE $\ell \leftarrow \ell -1$;
\ELSE
\STATE $c_{L-t-1} = 0$;
\ENDIF
\ENDFOR
\end{algorithmic}
\end{algorithm}

The $k$-bit information $\bvec{b}$ is recovered by decoding the codeword $\bvec{c}$ using Algorithm \ref{alg_decode}, where “$\gg$" stands for right bit-shift operator.

\begin{algorithm}[ht]
\caption{Decode $\bvec{c}$ into $\bvec{b}$}
\label{alg_decode}
\begin{algorithmic}[1]
 \renewcommand{\algorithmicrequire}{\textbf{Input:}}
 \renewcommand{\algorithmicensure}{\textbf{Output:}}
\REQUIRE $\alpha$, $L$, $\bvec{c}=(c_0, c_1, \ldots , c_{L-1})$, $c_t\in\{0,1\}$
\ENSURE  $\bvec{b}=(b_0, b_1, \ldots , b_{k-1})$, $b_t\in\{0,1\}$
\STATE $B \leftarrow 0$;
\STATE $\ell \leftarrow 0$;
\FOR{$t=0$ \TO $L-1$}
\IF{$c_t = 1$}
\STATE $\ell \leftarrow \ell +1$;
\STATE $B \leftarrow B + \displaystyle \binom{t}{\ell}$;
\ENDIF
\ENDFOR
\FOR{$t=0$ \TO $k-1$}
\STATE $b_t = B \pmod 2$
\STATE $B \leftarrow B \gg 1$;
\ENDFOR
\end{algorithmic}
\end{algorithm}

\section{Proposed Method}
\label{sec:prop}
The idea of proposed method is to encode the $k$-bit watermark $\bvec{b}$ into the codeword $\bvec{c}$ by using constant weight code before embedding operation.
Even when a pruning attack is done to round weight parameters $w_i$ with small value into 0, those elements are judged as bit 0 in the codeword, and hence, there is no effect on the received codeword.
As for bit 1, the corresponding weight parameters $w_i$ should be sufficiently large so that these are not cut off.

At the initialization of a given DNN model, $L$ weight parameters $\bvec{w}$ are selected from $N$ candidates according to a secret key. Then, an encoded watermark $\bvec{c}$ is embedded into $\bvec{w}$ under the constraint defined as follows:
\begin{itemize}
    \item If $c_i = 1$, then $|w_i| \ge T_1$; otherwise, $|w_i| \le T_0$, where $T_0$ and $T_1$ are thresholds satisfying $0 < T_0 < T_1$.
\end{itemize}
In the training process of DNN model, weight parameters are updated iteratively to be convergence into a local minimum.
The changes of the weights $\bvec{w}$ selected for embedding $\bvec{c}$ are only controlled by the above restriction during the training process in the proposed method.

\subsection{Embedding}
First, we encode a $k$-bit watermark $\bvec{b}$ into the codeword $\bvec{c}$ by using the Algorithm \ref{alg_encode}. Here, the parameters $\alpha$ and $L$ must be satisfy the following condition:
\begin{equation}
\label{eq_combi}
 2^k \le \binom{L}{\alpha}
= \frac{L!}{\alpha!(L-\alpha)!}< 2^{k+1}.
\end{equation}

At the embedding operation, the weight parameters $\bvec{w}$ selected from a DNN model are modified into $\bvec{w^\dagger}$ by using two threshold $T_1$ and $T_0$. 
\begin{equation}
\label{eq:embed}
w^\dagger_i = \left\{
\begin{array}{ll}
w_i & (c_i = 1) \cap (|w_i| \ge T_1) \\
sgn(w_i)\cdot T_1 & (c_i = 1) \cap (|w_i| < T_1)\\
w_i & (c_i = 0) \cap (|w_i| \le T_0) \\
sgn(w_i)\cdot T_0 & (c_i = 0) \cap (|w_i| > T_0)
\end{array}
\right.,
\end{equation}
where
\begin{equation}
    sgn(x) = \left\{
    \begin{array}{cc}
    1 & x \ge 0\\
    -1 & x < 0
    \end{array}
    \right..
\end{equation}
Eq.(\ref{eq:embed}) can be regarded as a constraint for executing the training process for the DNN model to embed  watermark.
Among huge number of $N$ candidates, we impose the constraint only to $L$ weight parameters selected for embedding.

\subsection{Detection}
At first, the weight parameters are selected from the same positions of a DNN model, which is denoted by $\bvec{w^\prime}$.
Then, the $\alpha$-th largest element is determined from $\bvec{w^\prime}$, and the codeword $\bvec{c^\prime}$ is constructed as follows:
\begin{equation}
 c^\prime_i = \left\{
	      \begin{array}{ll}
	       1 ~ & ~ \mbox{if} ~ |w^\prime_i| \ge \tilde{w}^\prime_{L-\alpha} \\
               0 ~ & ~ \mbox{otherwise}
              \end{array}
             \right. ,
\end{equation}
where $\bvec{\tilde{w}^\prime}=sort(|\bvec{w}^\prime|)$.
Finally, using the Algorithm \ref{alg_decode}, the watermark $\bvec{b^\prime}$ is reconstructed from the codeword $\bvec{c^\prime}$ as the result.

\subsection{Design of Two Thresholds}
Since the weight of codewords is constant, we select $\alpha$ largest elements from the $L$ elements in the weight parameter $\bvec{w^\prime}$ extracted from a given DNN model.
Though some weight parameters are cut off by the pruning attack, the values of such $\alpha$ elements are remained if the threshold $T_1$ is properly designed.

From to the definitions, we assume that the value of weight parameters in a DNN model is modeled as Gaussian distribution before and after the training of the model.
As the pruning attack cuts off $p$ weight parameters with small values, the absolute values of $\alpha$ elements in $\bvec{w^\dagger}$ should be more than them.
The statistical analysis of the distribution gives us the following inequality with respect to the threshold $T_1$.
\begin{eqnarray}
    R &\le& \frac{1}{\sqrt{2\pi\sigma^2}}\int_{-\infty}^{T_1}  \exp\Big(-\frac{x^2}{2\sigma^2}\Big)dx \\
    &=& 1- \frac{1}{\sqrt{2\pi\sigma^2}}\int_{T_1}^{\infty}  \exp\Big(-\frac{x^2}{2\sigma^2}\Big)dx \\
    &=& 1-Q\Big(\frac{T_1}{\sigma}\Big),
\end{eqnarray}
where $Q()$ is the Q-function. 
By using the inverse Q-function $Q^{-1}()$, the appropriate threshold $T_1$ can be calculated for a given pruning rate $R$.
\begin{equation}
    T_1 = \sigma Q^{-1}(1-R) 
\end{equation}

As for $T_0$, it is sufficient to give the condition $0 < T_0 < T_1$ if only pruning attack is assumed.
Considering the robustness against other attacks which will modify weight parameters in a watermarked model, we should give an appropriate margin $T_1-T_0$ by setting $T_0$.

\subsection{Considerations}
The usability of the proposed embedding method is confirmed from the studies of previous DNN watermarking methods.
For instance, the constraint given by Eq.(\ref{eq:embed}) can be applied for the embedding operations in \cite{uchida_ICMR17,nagai_IJMIR18, chen_ICMR19, rouhani_ASPLOS19}.
In the case of method \cite{Kuribayashi_IHMMSEC2021}, the embedding operation based on the constraint can be regarded as the initial assignment of weight parameters to a DNN model and the change of the weights at each epoch is corrected by iteratively performing the operation.

From the secrecy point of view, it is better to select small $\alpha$.
There are two possible approaches for attackers to cause a bit-flip in the codeword embedded into a DNN model.

One approach is to identify the elements those satisfying $|w_i|\ge T_1$ and to decrease their weight values.
Among $N(1-R)$ candidates of weight parameters $|w_i|\ge T_1$, identifying $\alpha$ values becomes difficult with the increase of $N$.
From the Definition \ref{def:scale}, the total number of weight parameters are much huge, and hence, this approach is difficult without seriously changing the weight parameters in a DNN model.

The other approach is to increase weight values at the selected weights those values are $|w_i| \le T_0$.
Again from the Definition \ref{def:scale}, it is extremely difficult to find such weights without a secret key.

\section{Numerical Examples}
\label{sec:example}
Table \ref{tab:example} enumerates some examples of parameters for constant weight code. For instance, when 64-bit watermark is encoded with $\alpha=10$, the length of its codeword becomes $L=393$. Then, the code can tolerate for the pruning attack with rate $R<0.9746$.

If the amount of watermark information is large, it is possible to divide it into small blocks and to embed each encoded block into selected weight parameters without overlapping.

\begin{table}
    \centering
    \caption{Numerical examples of parameters.}
    \label{tab:example}
    \begin{tabular}{|c|c|c|c|}\hline
    $k$ & $\alpha$ & $L$ & $\alpha/L$ \\\hline
    64 & 8 & 972 & 0.9918\\
       & 9 & 583 & 0.9846\\
       & 10 & 393 & 0.9746\\
       & 11 & 288 & 0.9618\\\hline
    128 & 16 & 1757 & 0.9909\\
        & 18 & 1063 & 0.9831\\
        & 20 & 722 & 0.9723\\
        & 22 & 533 & 0.9587\\\hline
    254 & 32 & 3307 & 0.9903\\
        & 36 & 2011 & 0.9821\\
        & 40 & 1373 & 0.9709\\
        & 43 & 1090 & 0.9606\\\hline
    512 & 63 & 6858 & 0.9908\\
        & 73 & 3693 & 0.9802\\
        & 79 & 2780 & 0.9716\\
        & 85 & 2196 & 0.9613\\\hline
    1024 & 127 & 12955 & 0.9902\\
         & 145 & 7443 & 0.9805\\
         & 159 & 5350 & 0.9703\\
         & 170 & 4323 & 0.9607\\\hline
    \end{tabular}
\end{table}

\section{Conclusions and Future Works}
\label{sec:conclude}
In this paper, we proposed a novel method for immunization of pruning attack in DNN watermarking by introducing the constant weight codes.
There are two thresholds $T_1$ and $T_0$ for giving a restriction of weight parameters to embed watermark.
Under the Gaussian assumption, $T_1$ can be calculated from a statistical analysis, while $T_0$ should be designed to consider the robustness against other possible attacks on the watermarked DNN model.

There are some studies about constant weight codes with error correcting capability~\cite{Gyorfi, Bitan, Etzion}.
The use of such codes is one of promising approaches to enhance the robustness against other attacks for DNN watermarking.

Many pruning algorithms have focused on compressing pre-trained models by removing weights whose absolute value is small, while an iterative magnitude pruning is proposed in \cite{Frankle_ICMR2019} that can find sparse sub-networks in randomly-initialized neural networks. One of our future works is to consider the robustness against such a sophisticated pruning algorithm.

\section*{Acknowledgment} 
This research was supported by the JSPS KAKENHI Grant Number 19K22846, JST SICORP Grant Number JPMJSC20C3, and  JST CREST Grant Number JPMJCR20D3, Japan.

\bibliographystyle{IEEEbib}
\bibliography{main}

\end{document}